\documentclass[amssymb,prd,aps,amsmath,twocolumn,superscriptaddress,nofootinbib]{revtex4}
\usepackage{pslatex}
\usepackage{graphicx}
\usepackage{psfrag}

\usepackage[usenames]{color}
\usepackage{color,graphicx,epsfig,amsmath,amssymb}

\newcommand{\beq}{\begin{equation}}
\newcommand{\eeq}{\end{equation}}
\newcommand{\ben}{\begin{eqnarray}}
\newcommand{\een}{\end{eqnarray}}
\newcommand{\bi}{\begin{itemize}}
\newcommand{\ei}{\end{itemize}}

\newcommand{\ghost}[1]{ }

\begin{document}

\title{Implication of the PAMELA antiproton data for dark matter indirect detection at LHC}

\author{C\'eline B\oe hm}
\affiliation{LAPTH, Universit\'e de Savoie, CNRS, BP110, F-74941 Annecy-le-Vieux Cedex, France}
\email{celine.boehm@cern.ch}

\author{Timur Delahaye}
\affiliation{LAPTH, Universit\'e de Savoie, CNRS, BP110, F-74941 Annecy-le-Vieux Cedex, France}
\affiliation{Dipartimento di Fisica Teorica, Universit\`a di Torino and INFN  Sezione Torino, via P. Giuria 1, I10125 Torino, Italy}
\email{timour.delahaye@lapp.in2p3.fr}

\author{Pierre Salati}
\affiliation{LAPTH, Universit\'e de Savoie, CNRS, BP110, F-74941 Annecy-le-Vieux Cedex, France}
\email{pierre.salati@lapp.in2p3.fr}

\author{Florian Staub}
\affiliation{Institut f\"ur Theoretische Physik und Astronomie, 
Universit\"at W\"urzburg, 97074 W\"urzburg, Germany}
\email{florian.staub@physik.uni-wuerzburg.de}

\author{Ritesh K. Singh}
\affiliation{Institut f\"ur Theoretische Physik und Astronomie, 
Universit\"at W\"urzburg, 97074 W\"urzburg, Germany}
\email{singh@physik.uni-wuerzburg.de}

\date{today}

\begin{abstract}
Since the PAMELA results on the ``anomalously'' high positron fraction and the lack
of antiproton excess in our Galaxy, there has been a tremendous number of studies 
advocating new types of dark matter, with larger couplings to electrons than to quarks.
This raises the question of the production of dark matter particles (and heavy associated coloured states) at LHC.
Here, we explore a very simple benchmark dark matter model and show that,
in spite of the agreement between the PAMELA antiproton measurements and the
expected astrophysical secondary background, there is room for large couplings
of a WIMP candidate to heavy quarks. Contrary to what could have been naively anticipated,
the PAMELA ${\bar{\rm p}}/{\rm p}$ measurements do not challenge dark matter model
building, as far as the quark sector is concerned. A quarkophillic species is
therefore not forbidden.Owing to these large couplings, one would expect that 
a new production channel opens up at the LHC, through quark--quark
and quark--gluon interactions. Alas, when the PDF of the quark is taken into account,
prospects for a copious production fade away.
%
%
%
\end{abstract}
\maketitle

\section{Introduction}
\label{sec:intro}

%
The new results of experiments such as PAMELA \cite{2009Natur.458..607A,2009PhRvL.102e1101A}, ATIC \cite{2008Natur.456..362C}, 
HESS \cite{2009arXiv0905.0105H}, FERMI \cite{2009PhRvL.102r1101A} have attracted
a lot of attention in the dark matter community during the last few months
\cite{Bergstrom:2008gr, Hooper:2008kv, Lattanzi:2008qa,2009NuPhB.813....1C} and somehow revived hopes
that cosmic ray data could finally shed light on the nature of dark matter.
In particular, the apparent conflict between the seeming excess in the positron
fraction and the lack of visible anomalies in the antiproton spectrum has encouraged
theoreticians to propose new dark matter models with very unusual properties.
For example, to fit PAMELA data, a large amount of dark matter models with very
large values of the annihilation cross section into leptons and small values of
the annihilation cross section into quarks have been proposed in the literature.
All these models rely on ``local'' enhancement mechanisms so that the candidate
abundance remains compatible with the measured dark matter relic density while,
simultaneously, leading to a very high amount of positrons in our galaxy. 
The complexity of the mechanisms involved illustrates the difficulty to fit the data 
and reflects the ongoing theoretical effort to explain the PAMELA positron excess.
%
However, this positron anomaly could be the signature of nearby pulsars
supplementing the interstellar medium with a radiation of electrons and positrons.
The possibility that spallation reactions take place during the acceleration of
primary cosmic rays has also been suggested.
%
%
In this Letter, we concentrate on the antiproton measurements which are subject
to neither speculation nor controversial interpretation. It has been shown indeed
that the PAMELA
${\bar{\rm p}}/{\rm p}$ ratio is compatible with a pure astrophysical
explanation. Antiprotons are generated inside the galactic disc and result from
the interaction of high energy cosmic ray protons and helium nuclei on the
interstellar material.
The lack of excess in antiprotons may suggest that the WIMP couplings to quarks must be small. 
However, here, we propose to find what is the maximal coupling that is allowed by the PAMELA antiproton data 
  to verify whether this assertion is true or not.

For simplicity, we base our analysis on a generic dark matter model where
the WIMP $\chi$ is directly coupled to a Standard Model quark $q$ and a heavy coloured
partner $F_{q}$. Inspired by supersymmetry, we shall focus this Letter on a Majorana WIMP so
$F_{q}$ is in fact a scalar (and would be equivalent to a squark in supersymmetry). 
Our approach is generic enough to be applied to other types of dark matter. 
However a complete survey of the various possibilities (including
scalar dark matter species $\chi$ and fermionic partner $F_{q}$) is  beyond the scope
of this work (although we anticipate that the results presented here are fairly general).

As we will show, the couplings that we obtain are in fact large enough to also have to 
address the question of the implications of the PAMELA antiproton data for the prospects of dark matter 
indirect detection at LHC.

In ``Standard'' supersymmetry, the heavy states $F_q$ are produced
mostly through gluon--gluon fusion proton-proton collisions. 
They are expected decay into missing energy
(the dark matter) and a jet (corresponding to the quark $q$) shortly after production.
Hence, this channel constitutes an important source for 
dark matter particles production at LHC.

However, if the direct dark matter couplings to quarks are larger than usually expected in ``standard'' 
supersymmetry, new channels could open up 
with $qq$, $q \bar{q}$ and $qg$ interactions, thus providing new possible signatures and increasing the discovery potential at LHC. 
Hence the importance of characterising the direct dark matter couplings to quarks in light of PAMELA antiproton data.

%
%
To constrain the dark matter characteristics (mass and pro\-perties of the particles
to which dark matter is coupled), we proceed as follows:
we first use in Sec.~\ref{timur} a realistic cosmic ray propagation code
to calculate the antiproton flux at the earth expected from conventional
spallations and dark matter annihilations. The latter depend on the
${\chi}-{q}-{F_{q}}$ left and right couplings $c_{L}^{q}$ and $c_{R}^{q}$
which we want to constrain from the PAMELA antiproton measurements.
We then consider a variety of annihilation channels and derive the maximal
cross sections allowed by the PAMELA antiproton data. Results are expressed
in units of the canonical thermal value of
$3 \times 10^{-26}$ cm$^{3}$ s$^{-1}$ and referred to as the boost factor
hereafter.
Because couplings to $u$, $d$ and $s$ quarks are already severely constrained
by direct detection, we focus on $c$ and $b$ quarks. With the help of a Monte
Carlo Markov Chain, we delineate in Sec.~\ref{celine} the region of parameter
space which saturates the upper bound on the boost.
Finally, in Sec.~\ref{LHC_results}, we determine the number of events associated
with
$qq \rightarrow FF$,
$q \bar{q} \rightarrow FF^*$ and
$q g \rightarrow F \chi$ production at LHC and discuss the propects for detecting
dark matter related signatures at LHC.
We conclude in Sec.~\ref{conclusion}.

%
\section{The PAMELA antiproton constraints}

\label{timur}

The antiproton flux at the earth arises from two sources. Conventional (secondary)
antiprotons are produced by the interactions of high energy cosmic ray protons and
helium nuclei with the gas of the galactic disc. The annihilation of hypothetical dark
matter species in the halo of the Milky Way generates additional (primary) antiprotons.
To compute both signals, we have considered the same propagation model as in \cite{2001ApJ...563..172D}
with the parameters best fitting the
boron to carbon ratio \cite{2001ApJ...555..585M}. As regards the secondary component, we have used the same local proton and helium nuclei 
fluxes as in \cite{Donato:2008jk} together with the radial distribution of SN remnants given by \cite{1992ApJ...390...96W} to 
retropropagate these fluxes all over the diffusive halo.

Concerning the primaries, we have considered a dark matter halo computed by \cite{vialactea} with
a local density of $\rho_\odot~=~0.3~\,~\rm{GeV \, cm^{-3}}$. 
The distance from the earth to the galactic
center is 8.5 kpc. The annihilation cross section is set to the conventional thermal value
mentioned above. We have determined the boost by which that value can be increased without
exceeding the PAMELA data. We have scanned the WIMP mass from 100 GeV to 1 TeV and considered
different annihilation channels. For each channel, the antiproton spectrum before propagation
has been calculated with the PYTHIA 6.4 program \cite{Sjostrand:2006za}. The results are summed
up in Tab.~\ref{tab:boosts}.

%
%
\begin{table}[ht] 
\footnotesize
	\centering
		\begin{tabular}{|c||c|c|c|c|c|c|c|c|c|c|} 
		\hline
	  & u $\bar{\text{u}}$ & d $\bar{\text{d}}$  & s $\bar{\text{s}}$ & c $\bar{\text{c}}$ & b $\bar{\text{b}}$  & g g & h h & Z$_0$ Z$_0$ & \tiny{W$^+$W$^-$} \\
  \hline
  \hline
		100 & 2 & 2 & 2 & 2 & 3 & 2 &  & & \\
		  \hline
		200 & 4 & 3 & 3 & 5 & 6  & 3 &  & 6 & 6 \\
		  \hline
		300 & 5 & 4 & 4 & 5 & 9 & 5 & 9 & 9 & 7 \\
		  \hline
		400 & 6 & 5  & 5 & 6 & 10 & 6 & 14 & 9 & 8 \\
		  \hline
		500 & 7 & 6  & 7 & 8 & 11 & 7 & 14 & 11& 9 \\
		  \hline
		600 & 8 & 8  & 8 & 9 &  12  & 7 & 15 & 12&11 \\
		  \hline
		700 & 10 & 9  & 10 & 11 & 13  & 8 & 16 & 1 4& 13 \\
		  \hline
		800 & 12 & 11 & 11 &  12 & 15 & 9 & 17 & 16 & 16 \\
		  \hline
		900 & 14 & 13  & 13 &  1 4& 17 & 10 & 19 & 19 & 18 \\
		  \hline
		1000 & 16 & 15 & 15 & 16 & 19  & 11 & 21 & 22 & 21\\
			\hline	
		\end{tabular}
	\caption{
	Maximum annihilation cross section allowed by the PAMELA antiproton measurements
	for various WIMP masses and different annihilation channels. This cross section
	is expressed in units of the canonical value of $3 \times 10^{-26}$ cm$^{3}$ s$^{-1}$.}
	\label{tab:boosts}
\end{table}
%

These constraints are quite severe. However, antiproton flux calculations suffer from a lot
of uncertainties and changing some of the choices we made may affect these results. As recalled
in \cite{1996PhR...267..195J} the local value of the dark matter density is not very well constrained
and in fact lies in $[0.2 ; 0.9]$ GeV cm$^{-3}$. The primary antiproton flux being proportional to the
square of the local density, one can divide all the results of Tab.~\ref{tab:boosts} by a factor ranging
from $\sim$ 0.4 to 9. The other uncertainties are summed up in Tab.~\ref{tab:uncertainties}.

%
%
\begin{table}[ht] 
\footnotesize
	\centering
		\begin{tabular}{|c||c|c|c|c|c|c|c|c|c|} 
		\hline
	  & ref & 1 & 2 & 3 & 4 & 5 & 6 & 7 \\
  \hline
  \hline
  		100 & 3 & 28  & 2 & 4 & 4 & 6 &  2 &2 \\
		  \hline
		200 & 6 & 60  & 3 & 8 & 6 & 11 & 3 & 4 \\
		  \hline
		300 & 9 & 90 & 4 & 13 & 10 & 18 & 5 & 5\\
		  \hline
		400 & 10 & 130 & 4 & 15 & 11 & 22 & 7 & 6\\
		  \hline
		500 & 11 & 150 & 4 & 16 & 12 & 23 & 9 & 6\\
		  \hline
		600 & 12 & 170 & 4 & 18 & 13 & 26 & 10 & 7 \\
		  \hline
		700 & 13 & 190  & 5 & 20 & 15  & 29 & 11 & 8\\
		  \hline
		800 & 15 & 220 & 6 & 23 & 17 & 33 & 12 & 9\\
		  \hline
		900 & 17 & 240 & 6 & 26 & 19 & 38 & 14 & 10 \\
		  \hline
		1000 & 19 & 270 & 7 & 29 & 21 & 42 & 16 & 11\\
			\hline	
		\end{tabular}
	\caption{
	Maximum boost (\textit{i.e.} annihilation cross--section in units of the canonical thermal value of
$3 \times 10^{-26}$ cm$^{3}$ s$^{-1}$) allowed by the PAMELA antiproton measurements for various WIMP masses
	in the case of annihilation into $b \bar{b}$ pairs when varying the parameters.
	Cases 1 and 2 correspond to extreme propagation parameters in agreement with B/C. 
	For case 2 the boost has been rounded to the closest decade. Cases 3, 4 and 5 correspond
	to various dark matter halo profiles~: NFW~\cite{nfw} (3), Moore~\cite{moore} (4) and
	an isothermal cored profile (5). Cases 6 and 7 correspond to alternative fits of the
	injection proton and helium nuclei spectra respectively proposed by \cite{2001ApJ...563..172D} and
	\cite{Shikaze:2006je}.}
	\label{tab:uncertainties}
\end{table}
%
%

%
One should notice that for most cases, the constraints come from the point of highest energy (61.2 GeV)
published by PAMELA. This point is the one that suffers the biggest statistical error and its systematic
error is unknown yet. Hence the results may change with future publication of new PAMELA data. Indeed
the current data correspond to only 500 days of data collection starting the 15th of June 2006 but the
satellite is still in orbit and should carry on taking data for at least few more months. Moreover,
the uncertainty related to the injection spectra should diminish as soon as absolute fluxes for protons
and antiprotons are published. Finally, new data are also expected from PAMELA for the boron to carbon
ratio that should help us limit the uncertainties on propagation parameters.

\section{The MCMC and the allowed region in parameter space}
\label{celine}

To find the region of parameter space that reproduces the boost factor as determined in the previous section, 
we perform a Markov Chain Monte Carlo search.
The free parameters of our benchmark model are a priori the dark matter mass,
the masses of the heavy scalar partners $F_{q}$, and their couplings
$c_{L}^{q}$ and $c_{R}^{q}$ to Standard Model quarks $q$.
Since $b$-quarks are a bit easier to tag (and direct detection experiments indicate that
the couplings to $u$, $d$ and $s$ quarks 
must be suppressed), we shall only focus on predictions associated with $b$ quarks.  
The only free parameters of interest are therefore the mass $m_{F_{b}}$ of the $b$~-~coloured state ${F_{b}}$,
the dark matter mass $m_{\chi}$ and its couplings to the $b$ quark, namely $c_{L}^{b}$ and $c_{R}^{b}$.
Our purpose is to explore this 4-D parameter space in order to delineate
the regions where the boost is close to the maximal value allowed by PAMELA.
The expression of the annihilation cross section is borrowed from \cite{Boehm:2003hm}.
The range of masses and couplings which we consider lies between $[100,2500]$ GeV and $[0.01,3]$
respectively. A larger upper limit on the couplings would induce a loss of perturbativity.

\begin{figure*}[h]
	\centering
		\includegraphics[width=15cm]{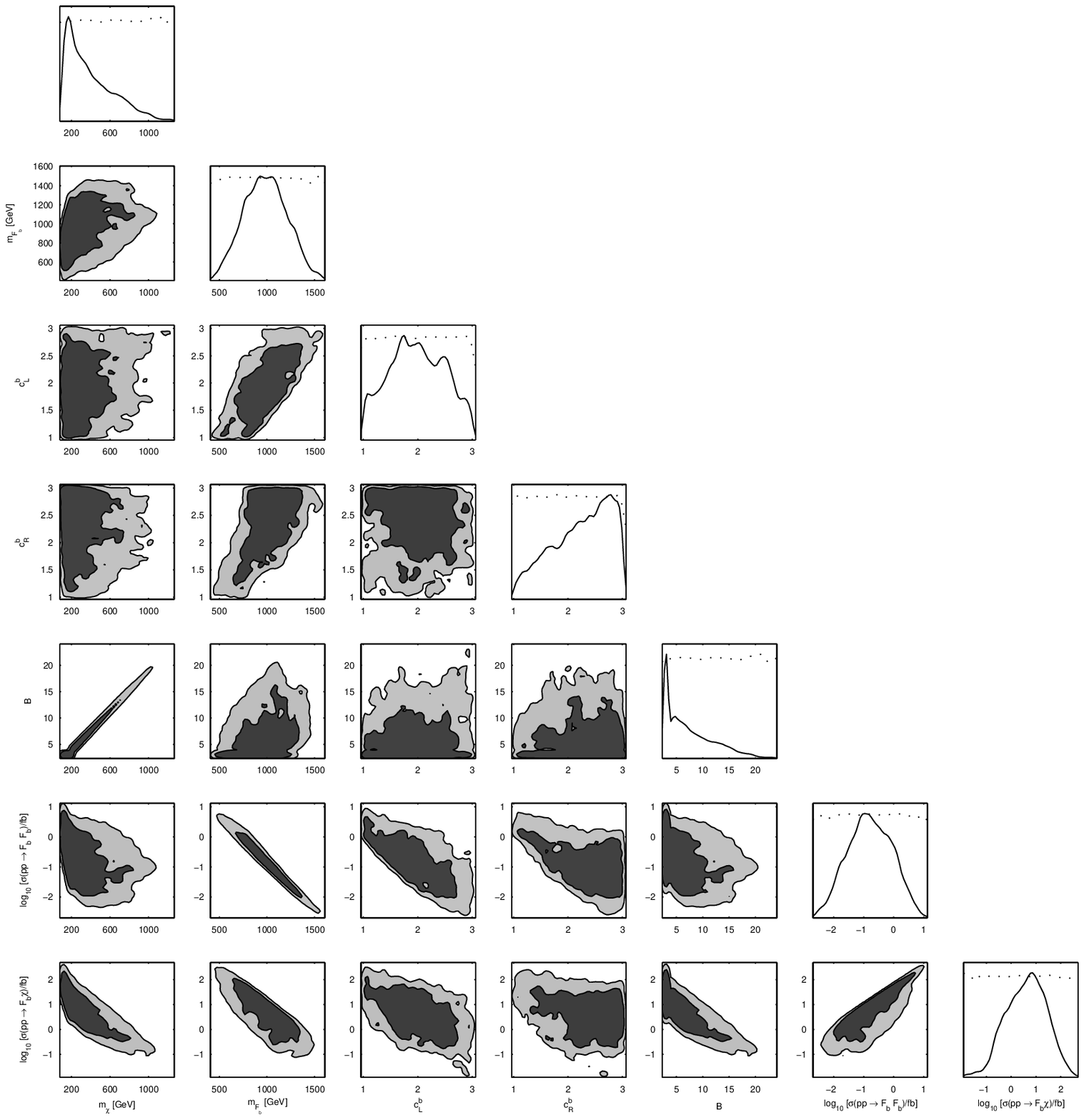}
		\caption{The dark matter parameter space is tunned to reproduce the boost factor obtained in Sec.~\ref{timur} for a $b \bar{b}$ pair.
		The central boost value is taken from Tab.~\ref{tab:boosts} while its uncertainty is taken to be $v_B=$0.1. Due to this very small boost uncertainty, one can see the very strong correlation between the boost value and the dark matter mass (see Tab.~\ref{tab:boosts}). The cross sections account for the $F_b F_b$, $\bar{F_b} F_b$, $F_b \bar{F_b}$ and $\bar{F_b} \bar{F_b}$ as well as the $F_b \chi$ and $\bar{F_b} \chi$ final states respectively. We used the getdist routine from the COSMOMC code \cite{Lewis:2002ah} to make this plot.}
		\label{fig:matlabdata_tri}
\end{figure*}

%
Translating the maximal boost $B_{\rm max}$ found in the previous section into
regions of the parameter space is tricky. For this purpose, we have built a Monte Carlo
Markov Chain which explores the parameter space and affects to each of its points
a likelihood depending on whether or not the boost is close to the maximal value allowed
by PAMELA. This method allows to rapidly delineate the interesting regions.
The likelihood is defined as a Gaussian centered on $B_{\rm max}$ with width
$v_{B} = 0.01$.
Results are displayed in the correlation plots of Fig.~\ref{fig:matlabdata_tri}. 
The first four rows correspond to the parameters $m_{\chi}$, $m_{F_{b}}$, $c_L^b$ and $c_R^b$
of the model while the fifth one is the reconstruction of the boost factor.
Notice that the first plot of this row corresponds (as it should) to the column
$b \bar{b}$ of Tab.~\ref{tab:boosts}.

The panel in the very first line represents the dark matter mass while the last panel of the second, third and fourth rows stands for $m_{F_b}$, $c_L^b$ and $c_R^b$ respectively. 
The first panel of the second line is a correlation plot between 
$m_{F_b}$ and $m_{\chi}$. The first (second) plot in the third line features the correlation between $c_{L}^b$ and 
$m_{\chi}$ ( $c_{L}^b$ and $m_{F_b}$) and so on.
The dotted lines in all the plots feature the prior distribution while the solid line is the posterior 
distribution, {\it i.e.}, the values of the parameters that are found by the Markov chain.
The chain has explored enough points since the prior distributions are indeed rather flat for the four parameters
while it matches a Gaussian distribution for the boost.
We can also check that the contraint of dark matter stability ($m_{F_b} > m_{\chi}$) is correctly
reproduced in the ($m_{F_b} , m_{\chi}$) plot of the second row.

\begin{figure*}[h]
	\centering
		\includegraphics[width=15cm]{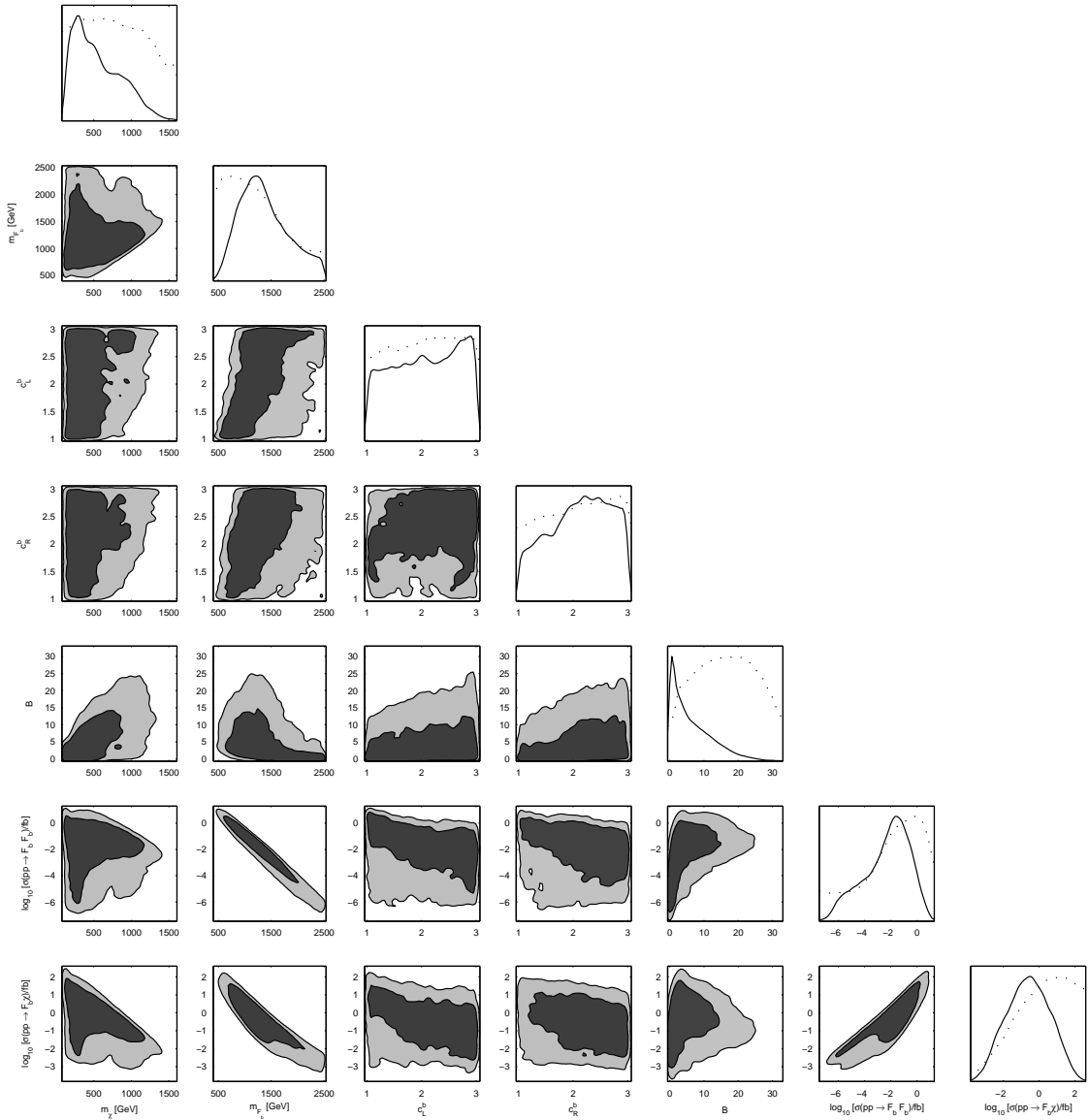}
		\caption{The dark matter parameter space is tunned to reproduce the boost factor obtained in Sec.~\ref{timur} for a $b \bar{b}$ pair.
		The central boost value is taken from Tab.~\ref{tab:boosts}. Its uncertainty is estimated from the smallest difference between the boost value displayed in the last five columns of Tab.~\ref{tab:uncertainties} and the central value of reference. Whatever the dark matter mass, the uncertainty $v_B$ thus obtained is much larger than in Fig.~\ref{fig:matlabdata_tri} and one loses the strong correlation between the value of the boost and the dark matter mass. Such ``large'' values of $v_B$ relaxes the constraints previously obtained and opens up the parameter space. As a result, large coupling values ($c_{l,r}^b \in [1,3]$) allow for larger $m_{F_b}$ values which translate into smaller production cross sections at LHC. Here we use $\sqrt{s}$=10 TeV.}
		\label{fig:matlabdata_tri_large}
\end{figure*}

Points of the parameter space are selected according to how the boost
is close to $B_{\rm max}$. The MCMC considers a priori equally light and
heavy WIMPs. But the heavier the WIMP, the larger the value of $B_{\rm max}$
as is clear in Tab.~\ref{tab:boosts} as well as in the first plot of the fifth
row of Fig.~\ref{fig:matlabdata_tri}. On the other hand, a heavy WIMP means
an even heavier coloured partner $F_{b}$ and since the annihilation cross
section is a decreasing function of $m_{F_{b}}$, a large boost is only
recovered for very large values of the couplings $c_L^b$ and $c_R^b$.
Hence, we expect the MCMC to select large values for
these couplings. 
This is confirmed by our results since both left and right handed couplings can reach two or
three units (with 90$\%$ and 68$\%$ CL respectively). If we consider a larger
variance $v_{B}$ as in Fig.~\ref{fig:matlabdata_tri_large}, the trend is the
same. Although slightly smaller values of the boost $B$ can be in principle
achieved with simultaneously large values of the couplings and mass $m_{F_b}$,
the latter would have to be larger than the 2.5 TeV limit which we set. Hence,
in this case, our Monte Carlo chooses a large value of one of the coupling (e.g. $c_L^b$)
together with a smaller value of the other coupling (e.g. $c_R^b$) and a large value
of $m_{F_b}$ (still within our limits).
Our main conclusion is that the PAMELA antiproton observations do not
preclude large values for $c_L^b$ and $c_R^b$. Our naive benchmark model
provides therefore a simple example where the WIMP can be quarkophilic
and yet satisfy the PAMELA constraints.



Note that the correlation plots displayed in Fig.~\ref{fig:matlabdata_tri} are obtained by saturating
the upper limit on the boost obtained in section \ref{timur}. The parameters found in these plots
therefore correspond to a large annihilation cross section into quarks, which translates 
into a relic density smaller than $\Omega_{dm}~h^2~=~0.1099~\pm~0.0062$ \cite{Dunkley:2008ie}. 
Since our model simply relies on WIMP couplings to quarks (we do not invoke the Sommerfeld mechanism
for example nor annihilations into heavy gauge bosons), this means that we have delineated the region
of the parameter space where WIMPs have the ``wrong'' relic density. Discovering heavy coloured states
at LHC decaying into jet + missing energy and with properties matching this region of the parameter
space would therefore jeopardise conventional dark matter scenarios.

\section{Heavy states production at LHC}
\label{LHC_results}

Since the WIMP couplings to b quarks can be very large, the question of $F_q$ production at LHC from  quark-quark collisions arises. We are interested in fact in  $b b \rightarrow F_b F_b$ (which may happen if dark matter is a real scalar or, in our case, a Majorana particle), $b \bar{b} \rightarrow F_b \bar{F_b}$ and $q g \rightarrow F \chi$. Interestingly enough, the  $qq \rightarrow FF$ and $qg \rightarrow F \chi$ cross sections can reach up a few pb or a few hundred $fb$ respectively for large couplings. 

Such large cross sections seem rather encouraging. However, to get realistic estimates, one has to take into account the parton distribution function (pdf) associated with b-quarks and gluons. By focusing on $b$, we have definitely considered a case where there is a very large suppression due to the pdf \cite{Whalley:2005nh}. However, the pdf for $c$
 quarks is also suppressed and the boost factor is not so large. Hence $b$ quarks are more or less representative of what is to be \textit{at most} expected at LHC for these types of $F_b$ production processes, given PAMELA antiproton measurements. 
 
At large $x$ and $Q^2$, convoluting our cross section with the parton distribution function leads to a large suppression of the number of events.  We predict less than ten events for $pp \rightarrow F_b F_b$ (for a integrated luminosity of $fb^{-1}$) in the region where the $bb \rightarrow F_bF_b$ process is supposed to be maximal (i.e. for large couplings) \cite{Belanger:2007zz}. The $pp \rightarrow F\chi$ cross section is a bit less suppressed because it involves gluon. However, it is still very small (it can reach at most ten $fb$, assuming a very small uncertainty on the value of the boost factor). Assuming a greater uncertainty on the boost value does not help. The two cross sections become even smaller since, in this case,  larger $m_{F_b}$ values are preferred when the dark matter couplings to quarks is greater than unity. 
Both cases lead to a too small number of events to be detected via the decay (jet$+$missing energy) of the heavy coloured states $F_b$. 
Indeed, they would correspond to pure hadronic final states which are extremely difficult to exploit. 
 
\vspace{1cm}

\section{Conclusion \label{conclusion}} 
In this letter, 
we have investigated the constraints that PAMELA set on the WIMP couplings to quarks. We found that coupling values larger than the 
strong coupling constant could be allowed by PAMELA antiproton measurements. This encouraged us to look at the production of heavy coloured final states from proton proton collisions.   We focus in particular on collisions producing $F_b$ particles (and therefore $b+\chi$) in the final state, since they might be easi\-er to tag than $c$-quarks. Also we ignore couplings to $u,d,s$ quarks since dark matter direct detection constraints indicate that these  must be suppressed.

We find that, despite the large values of the WIMPs couplings to quarks, there is very little hope to produce a large number of coloured states $F_b$ through $bb$, $b\bar{b}$,{bg} process in $pp$ collisions because of the b quark pdf suppression (less than a few 100 events at LHC for most of the parameter space under consideration and a luminosity of $1 fb^{-1}$). 
Also, such a final state would be purely hadronic and would be difficult to disentangle from background.
Hence, the best channel to constrain the WIMP (direct) couplings to quarks in the absence of other signatures 
may be, indeed, heavy coloured states production via gluon fusion and their decay into jets plus missing energy.

Although there have been many studies advocating leptophilic dark matter particles to fit the PAMELA positron excess, our analysis shows  that the PAMELA antiproton measurement also allows for ``quarkphilic'' dark matter in principle. This property may hold whether dark matter is at the origin of the positron excess or not. To fit both PAMELA antiproton data and positron excess, it therefore seems likely that only the WIMP couplings to gauge boson (the $W$ in particular) may have to be suppressed. Not only are large WIMPs couplings to heavy quarks allowed by PAMELA antiproton data but they may also be difficult to constrain using LHC data since their effect on the cross section can be compensated by very large $F_q$ masses and is suppressed by heavy quark pdf folding.
 
This actually constitutes the key point of our analysis and suggests that the dark matter production channel in this simple model will also rely on $F_q$ production through gluon gluon fusion. Hence, such ``quarphilic'' models may be difficult to disentangle from standard supersymmetry, unless one can measure the couplings very accurately. In any case, this analysis shows that even though leptophilic dark matter particles may fit the PAMELA positron excess data, this does not imply that the dark matter couplings to quarks must be suppressed. 
On the other hand,  the strength of the dark matter-quark interactions will remain relatively weak owing to the exchange of heavy messengers $F_q$.

\vspace{2cm}
\begin{acknowledgments}
We are grateful to Peter Skands who was very kind to help us in dealing with Pythia code,  J. Hamann and  J. Lesgourgues for their help with getdist routine. We would like to thank F. Arleo, F. Boudjema, J. Idarraga and W. Porod for fruitful discussions. T.D. is grateful for financial support from the Rh\^one-Alpes region (Explora'Doc) and from the International Doctoral school in AstroParticle Physics (IDAPP). F.S. is supported by the DFG Graduiertenkolleg GRK-1147 while the work of R.K.S. is supported by the German BMBF under contract 05HT6WWA.
We would like to thank the referee for his/her suggestions.
\end{acknowledgments}


\end{document}